\documentclass[12pt]{article}
\usepackage{amssymb,amsmath,amsfonts,amsthm,eurosym,graphicx,color,setspace,sectsty,comment,footmisc,pdflscape,subfigure,array}
\usepackage{subcaption}
\usepackage[comma,authoryear]{natbib}
\usepackage[colorlinks=true,urlcolor=blue,citecolor=blue,linkcolor=blue,bookmarks=true]{hyperref}
\usepackage[font=small,skip=0pt]{caption}
\usepackage{cmap} 
\usepackage{mathtext}
\usepackage{rotating}
\usepackage{array,makecell,booktabs,multirow}
\newcolumntype{C}[1]{>{\centering\arraybackslash}p{#1}} 
\usepackage{booktabs,multirow,makecell,array}
\usepackage{bbm}
\usepackage{adforn} 
\usepackage{alphalph}
\usepackage[titletoc]{appendix}
\usepackage{commath}
\usepackage{datetime}
\usepackage{placeins}
\usepackage{footnote}
\usepackage{graphicx}
\usepackage{subcaption}
\usepackage{booktabs}
\usepackage{pdfpages}
\usepackage{fancyhdr}
\usepackage{mathrsfs}
\usepackage{enumerate}
\usepackage{adjustbox}
\usepackage{booktabs}
\usepackage{epstopdf}
\usepackage{longtable}
\usepackage{multirow}
\usepackage{ragged2e}
\usepackage{tabularx}
\usepackage[para, flushleft]{threeparttablex}
\usepackage{float}
\usepackage{wrapfig} 
\usepackage[T2A]{fontenc} 
\usepackage[utf8]{inputenc} 
\usepackage[english]{babel}
\usepackage{pgf,tikz} 
\usepackage{pgfplots}
\usepackage{pgfplotstable}
\usetikzlibrary{shapes}
\usetikzlibrary{arrows, calc}
\usepackage{titlesec}
\usepackage{authblk}
\usepackage{indentfirst}
\usepackage{csquotes}
\usepackage[left=2cm,right=2cm, top=2cm,bottom=2cm,bindingoffset=0cm]{geometry}
\usepackage{url}
\usepackage{subcaption}
\usepackage{graphicx}
\usepackage{caption}
\usepackage{rotating}
\usepackage{xcolor}
\usepackage{array,booktabs,xurl}
\newcolumntype{Y}{>{\raggedright\arraybackslash}p{4.9cm}} 

\begin{document}

\title{Not Yet: Humans Outperform LLMs in \\ a Colonel Blotto Tournament}

\date{}

\author[1,2,*]{Dmitry Dagaev}
\author[1]{Egor Ivanov}
\author[1]{Petr Parshakov}
\author[3,4,5,6]{Alexey Savvateev}
\author[1]{Gleb Vasiliev}

\affil[1]{HSE University}
\affil[2]{New Economic School}
\affil[3]{Central Economic Mathematical Institute, Russian Academy of Sciences}
\affil[4]{Adyghe State University}
\affil[5]{Moscow Institute of Physics and Technology}
\affil[6]{Innopolis University}
\affil[*]{Corresponding author. Email: ddagaev@gmail.com. Address: 121205, Russia, Moscow, Innovation Centre Skolkovo, Nobel street, 3.}

\maketitle

\begin{abstract}

The emergence of large language models (LLMs) has spurred economists to study how humans and LLMs behave in strategic settings. We organized a series of round-robin tournaments in the Colonel Blotto game. This game attracts game theorists' attention due to high-dimensional action space and the absence of pure strategy Nash equilibria. In the first tournament, more than 200 human participants competed against one another. In the second tournament, several popular LLMs were invited to submit strategies. In the third tournament, we matched the number of LLM strategies to the number submitted by humans. We find that humans more often employ better-calibrated intermediate-level allocation heuristics and outperform the simpler, more stereotyped strategies submitted by LLMs. Strategic sophistication is key to success if and only if the necessary level of reasoning depth is reached, while lower and higher levels of reasoning offer no clear advantage over the primitive strategies. Among humans, field of study weakly predicts success: participants with STEM backgrounds perform better in the first tournament. Surprisingly, humans almost do not adjust their strategies across tournaments with different sets of opponents. This result suggests that humans base their choices primarily on the game’s rules rather than on the identity of their opponents, treating LLMs much like human competitors.
\end{abstract}

~~~~\textbf{Keywords}: Colonel Blotto, electoral college, tournament, $k$-level reasoning, AI, LLM

~~~~\textbf{JEL codes}: C72, C99, D91

\doublespacing
\newpage
\section{Introduction}

The rapid diffusion of large language models (LLMs) has revived several classic questions in behavioral and experimental economics. First, to what extent can artificial agents replicate, predict, or even substitute human decision-making in strategic environments? Second, how interaction between humans and LLMs differs from the interaction without artificial intelligence (AI)? On the one hand, LLM-based systems are increasingly deployed in settings where agents must interpret natural language, form beliefs, and interact strategically with other agents or humans. Their appeal is straightforward: for many codifiable, high-volume, or knowledge-intensive tasks, LLMs can be cost-competitive and productivity-enhancing, which has intensified interest in automation and its labor-market consequences \citep{Acemoglu2018}. On the other hand, LLMs are trained primarily to generate plausible text, not to optimize payoffs or to implement equilibrium reasoning. Hence, even when LLM outputs appear ``human-like,'' it remains unclear whether they generalize across incentive structures, strategic complexity, and informational environments.

A fast-growing literature has begun to evaluate LLMs as experimental participants, as simulated economic agents, and as building blocks for generative agent societies. A prominent line of work argues that LLMs can be treated as implicit computational models of human behavior, a ``Homo silicus'', that can be endowed with preferences and constraints and then studied in counterfactual environments \citep{Horton2023}. At the same time, evidence is accumulating that LLM behavior is highly sensitive to framing, prompts, model-specific safeguards, and the strategic structure of the task. For example, canonical laboratory games often reveal systematic deviations: LLMs may display unusually strong fairness and cooperation motives in one-shot settings such as the Dictator game \citep{Brookins2023}, while in preference elicitation tasks they can exhibit impatience and language-dependent anomalies \citep{GoliSingh2024}. In strategic depth tests, recent work cautions against using LLMs as human surrogates: in the 11--20 money request game, most models fail to reproduce human distributions of reasoning depth across many conditions \citep{Gao2025}. In repeated interactions, LLMs can perform well in self-interested repeated games but struggle in environments that require coordination \citep{Akata2025}.  \cite{alekseenko2025strategizing} contributed to this agenda by comparing human play and LLM play in a Keynesian beauty contest, documenting systematic differences in strategic sophistication with LLMs playing strategies closer to Nash equilibrium but failing to exploit dominant strategies in a game with two players. Beyond ``LLMs as subjects,'' benchmarks and arenas for agentic evaluation have also emerged, including economics-focused platforms designed to compare models in stylized economic tasks and competitive environments \citep{Guo2024}.

The present paper extends the comparison of humans and LLMs to a markedly different strategic environment, a Colonel Blotto game. The classical Colonel Blotto game is a two-player zero-sum resource-allocation contest. Each player simultaneously allocates a fixed budget of resources (e.g., troops, campaign effort, or expenditures) across $n$ independent battlefields (or districts). Payoffs are determined battlefield by battlefield: the player allocating more resources to a given battlefield wins that battlefield and receives its value (often normalized to 1, or weighted to reflect heterogeneous battlefield importance), while ties split the value. The player with the larger total value of won battlefields wins the game. 

The Colonel Blotto game has a long intellectual history. The first formulation dates back to \citet{borel1921theorie}. For $n=3$ and the case of symmetric players, Nash equilibria in mixed strategies were found in \cite{borel1938application}. \cite{gross1950continuous} solve the game with symmetric players for Nash equilibria in mixed strategies for arbitrary $n$. Also, \cite{gross1950continuous} found Nash equilibria in mixed strategies for $n=2$ and asymmetric players. It took more than 50 years to obtain the complete description of all Nash equilibria in mixed strategies for the case of asymmetric players and arbitrary $n$, when \cite{roberson2006colonel} solved the problem.

The Colonel Blotto game has a rich set of economic interpretations: it is closely related to distributive politics and electoral competition \citep{laslier2002distributive, washburn2013or}, R\&D races \citep{golman2009general}, sports \citep{rehsmann2023sumo}, and broader models of competitive allocation under scarcity. In these applications, politicians, investors, and coaches decide how to allocate the limited budget of time, money, or effort across multiple tasks.

Experimental research has documented interesting patterns in strategies used by players. The large set of possible strategies in the Colonel Blotto game leads to thinking in terms of strategies' features rather than in terms of strategies itself. Level $k$ reasoning manifests itself in such dimensions as deviation from the uniform distribution of resources and using round numbers \citep{arad2012multi}. In asymmetric games, players with lower budget of resources tend to focus on a particular subset of battlefields leading some of the battlefields empty \citep{avrahami2009weak}. Mixed evidence on equilibrium play in Colonel Blotto game was found in \citet{chowdhury2013experimental}.

Relative to prisoner's dilemma, beauty contests, and many canonical $2\times 2$ games, Blotto contests feature (i) high-dimensional action spaces, (ii) no dominant strategies, (iii) equilibria that typically require mixed strategies, and (iv) careful reasoning about opponents. The high-dimensional allocation problem creates many ``reasonable-sounding'' heuristics (e.g., proportionality rules or focal allocations) whose performance depends critically on what others do, making the environment informative about the depth and adaptability of reasoning. The absence of dominant strategies or pure-strategy Nash equilibrium requires correctly anticipating the distribution of opponents’ allocations rather than best-responding to a single deterministic plan. These characteristics make Colonel Blotto game a natural test for the strategic competence of artificial agents and a meaningful benchmark for comparing strategic reasoning across humans and LLMs.

Motivated by these considerations, we organized a series of round-robin tournaments in a Blotto environment with more than 200 human participants and a set of popular LLMs that were asked to submit strategies. The design consists of three tournaments. The first tournament is purely human: participants compete against one another. The second tournament introduces LLM-submitted strategies into the opponent pool. The third tournament equalizes the number of LLM strategies and human strategies, ensuring that performance comparisons are not mechanically driven by the relative mass of artificial strategies in the population. This design lets us separate (i) differences in strategic sophistication between humans and LLMs from (ii) composition effects arising from changes in the opponent set.

Our main findings are fourfold. First, humans systematically outperform the strategies submitted by LLMs in the tournament environment. Second, the performance gap is consistent with humans relying more often on better-calibrated intermediate-level heuristics, while many LLM strategies resemble simpler and more stereotyped allocation templates. Third, within the human sample, field of training weakly predicts success: participants with STEM backgrounds perform better in the first tournament, while gender, education, and employment status do not robustly predict outcomes. Fourth, despite strong theoretical incentives to condition play on the opponent pool in a game without dominant strategies, we find little evidence that humans adjust their strategies across tournaments with different opponent compositions. This suggests that, in this environment, humans may anchor on the rules of the game rather than on the identity of opponents, treating LLM strategies much like human strategies. Overall, our results add to the emerging evidence that LLM behavior in strategic settings is heterogeneous and context-dependent, and that performance in complex contests can reveal limitations that may be masked in simpler games or in prompt-sensitive vignette tasks.

The rest of the paper is organized as follows. Section 2 presents the design of our experiment. In Section 3, we describe the properties of the dataset. Section 4 contains results of the tournaments. Section 5 models the relationship between reasoning and performance. Section 6 discusses the results, and Section 7 concludes.

\section{Tournament Design}

We conducted three round-robin tournaments in the same strategic environment, the \textit{Electoral Race} game (a discrete Colonel Blotto variant). Each tournament consists of a set of submitted strategies. Within a tournament, every strategy plays exactly one match against every other strategy (round-robin). A match is a single play of the \textit{Electoral Race} game described below. Tournament outcomes are determined by the total number of match wins, with ties counted as half a win. When two or more strategies obtain the same tournament score, ties are broken by the time of submission: earlier submissions receive a higher rank.

The \textit{Electoral Race} game is a two-player zero-sum contest with a discrete allocation decision across nine states. Two candidates for president of a fictional country simultaneously and independently allocate a fixed campaign budget of 100 campaign trips across nine states, indexed by $s\in\{A,B,C,D,E,F,G,H,I\}$. A (pure) strategy is an integer vector
\[
x=(x_A,\ldots,x_I)\in\{0,1,\ldots,100\}^9
\quad \text{such that} \quad \sum_{s} x_s \leq 100,
\]
where $x_s$ denotes the number of trips allocated to state $s$. Given two strategies $x$ and $y$, the outcome in state $s$ is determined by comparing $x_s$ and $y_s$. The player with the larger allocation wins the state and receives 1 electoral vote; if $x_s=y_s$, the state is tied and each player receives 0.5 electoral votes. Total electoral votes are the sum across states. The match winner is the candidate who obtains strictly more electoral votes. The match payoff is 1 for the winner and 0 for the loser; if total electoral votes are equal, the match is a tie and each strategy receives 0.5.

The three tournaments differ only in the composition of the opponent pool. Tournament~1 includes only human-submitted strategies: each participant submitted one strategy, and all human strategies played one another in a round-robin format. Tournament~2 includes human strategies and a set of strategies submitted by popular large language models (LLMs), with each participating LLM submitting one strategy. Tournament~3 again includes human and LLM strategies, but in this tournament we matched the number of LLM strategies to the number of strategies submitted by humans. This design choice controls for the ``mass'' of LLM strategies in the tournament population. Participants were allowed to submit different strategies across tournaments.

Participation in Tournaments 1--3 was open to volunteers aged 18 or older. Each participant could submit at most one strategy per tournament. Strategies were submitted electronically by a common deadline (October~20,~2025, 23:59 Moscow time) which was further extended till (November~2,~2025, 23:59 Moscow time). All valid submitted strategies were included in the corresponding tournament.

Let $W_i$ denote the number of match wins achieved by strategy $i$ in a given tournament, and let $T_i$ denote the number of tied matches. The tournament score of strategy $i$ is defined as
\[
S_i = W_i + \tfrac{1}{2}T_i,
\]
and strategies are ranked by $S_i$ in descending order, with submission time used as a tie-breaker. 

In addition to tournament-specific rankings, we constructed an aggregate leaderboard based on performance in Tournaments~1--3. In each tournament, the top 100 ranked strategies (including LLM strategies) earned points toward the aggregate leaderboard. Points were assigned as follows: 200 points for rank 1, 180 for rank 2, 165 for rank 3, 150 for rank 4, 140 for rank 5, 130 for rank 6, 120 for rank 7, 110 for rank 8, 100 for rank 9, and for ranks $X\in\{10,\ldots,100\}$, $(101-X)$ points. If multiple strategies tied for a set of ranks, they split equally the total points assigned to those ranks. The aggregate rank is determined by the sum of points across Tournaments~1--3 with time of strategy submission serving as a tie-breaker.

Prizes were offered to incentivize participation, including awards for top-performing human strategies in each tournament and additional awards based on the aggregate leaderboard. Namely, top 10 authors in each tournament and additional top 10 performers in the general ranking received a sci-pop book co-authored and signed by one of the authors of this experiment. The winner of the aggregate leaderboard was offered a lunch with any of the experiment authors. 

The above mentioned regulations were announced in the advertisement of the experiment. Promotion of the tournaments was organized through Youtube and social networks of the authors. The authors did not directly advertise the experiment to students who presumably were familiar with the Colonel Blotto game (some of the coauthors of this paper discuss the Colonel Blotto game at Microeconomics or Game Theory classes. These cohorts of students did not receive any direct invitations to participate in the experiment).

All analyses were conducted on anonymized data: submissions were processed without personal identifiers except emails, and names were requested only from prize recipients for delivery purposes. In Appendix A, we provide the form that all participants had to fulfill when submitting a strategy. 

Submission of a strategy constituted consent to the tournament rules.

\section{Data}

Overall, 222 participants submitted their strategies. 
All submissions were checked prior to inclusion in the tournaments. 
Strategies that violated the rules of the game were excluded from the analysis. The most common reason for exclusion was exceeding the total budget of 100 campaign trips; a small number of submissions were also excluded for other formal inconsistencies. Duplicates were also removed. After excluding invalid submissions and removing duplicate entries, the final number of valid strategies admitted to the tournaments was 205 in Tournament~1, 207 in Tournament~2, and 207 in Tournament~3. In total, there are 215 players with at least 1 valid strategy.

In addition to the submitted strategies, we collected background information on the authors of the strategies. 
The survey included age, gender, education, current occupation, field of knowledge, a short free-text description of the participant's decision-making principle, and an email address for organizational purposes. 
The full survey instrument and the original response options are reported in the Appendix A.

\begin{table}[!htbp]\centering
\caption{Sample characteristics: sex, education, employment status, and field of study}
\label{tab:sample_characteristics}
\small
\begin{tabular}{p{3cm}p{2.3cm}p{7cm}rr}
\toprule
Variable & Category & Initial category & N & \% \\
\midrule
Sex & Female & Female & 44 & 20.5 \\
 & Male & Male & 171 & 79.5 \\
\addlinespace
Education & Secondary & Incomplete secondary & 9 & 4.2 \\
 &  & Secondary & 10 & 4.7 \\
 & Higher & Incomplete higher & 75 & 34.9 \\
 &  & Higher & 105 & 48.8 \\
 & Doctoral degree & Doctoral degree & 16 & 7.4 \\
\addlinespace
Employment Status & Not working & Temporarily not working (job search, parental leave, etc.) & 11 & 5.1 \\
 &  & Retired & 3 & 1.4 \\
 &  & Other & 6 & 2.8 \\
 & Student & Studying (school student, university student, PhD student) & 100 & 46.5 \\
 & Working & Employed (in an organization) & 70 & 32.6 \\
 &  & Self-employed / freelancer & 17 & 7.9 \\
 &  & Entrepreneur / business owner & 8 & 3.7 \\
\addlinespace
Field of Study & Economics and Management & Economics, business and management & 70 & 32.6 \\
 & STEM & Mathematics & 35 & 16.3 \\
 &  & Computer science and technical sciences & 56 & 26.0 \\
 &  & Natural sciences (physics, chemistry, biology, etc.) & 25 & 11.6 \\
  & Humanities, Social Sciences, and other & Humanities and arts & 7 & 3.3 \\
 &  & Social sciences (sociology, psychology, political science, etc.) & 10 & 4.7 \\
  &  & Other & 12 & 5.6 \\
\bottomrule
\end{tabular}
\end{table}

Table~\ref{tab:sample_characteristics} reports the main categorical characteristics of the sample. 
The sample is predominantly male: 79.5\% of participants are men, while 20.5\% are women. In terms of education, the vast majority of respondents have higher education or are currently studying at the university (83.7\%), whereas 8.8\% have secondary education and 7.4\% hold a doctoral degree. Regarding employment status, the sample is relatively balanced between students and employed respondents: 46.5\% are students and 44.2\% are working, while only 9.3\% report not working.

To reduce sparsity and facilitate interpretation, several survey variables were aggregated into broader categories. 
The original education variable, reported in the Appendix, contained multiple levels reflecting different stages of schooling and higher education. 
Respondents indicating \textit{Incomplete secondary} or \textit{Secondary} education were grouped into the category \textit{Secondary}. 
Those reporting \textit{Incomplete higher} or \textit{Higher} education were classified as \textit{Higher}. 
Respondents indicating a \textit{Doctoral degree} were assigned to the category \textit{Doctoral degree}.

Current occupation was also recoded into broader groups. 
Using the original categories reported in the Appendix, respondents indicating \textit{Studying (school student, university student, PhD student)} were classified as \textit{Student}. 
Those who selected \textit{Employed (in an organization)}, \textit{Self-employed / freelancer}, or \textit{Entrepreneur / business owner} were grouped into the category \textit{Working}. 
The remaining categories --- \textit{Temporarily not working}, \textit{Retired}, and \textit{Other} --- were combined into \textit{Not working}. 
In the empirical analysis, \textit{Not working} serves as the reference category.

Finally, respondents' fields of knowledge were aggregated into three broad domains. 
Based on the original survey categories listed in the Appendix, individuals who reported \textit{Mathematics}, \textit{Computer science and technical sciences}, or \textit{Natural sciences} were grouped into the \textit{STEM} category. 
Respondents indicating \textit{Economics, business and management} were classified as \textit{Economics and Management}. 
All other respondents were assigned to the reference category \textit{Humanities, Social Sciences, and other}.

\begin{figure}[htbp]
\centering
\includegraphics[width=0.65\linewidth]{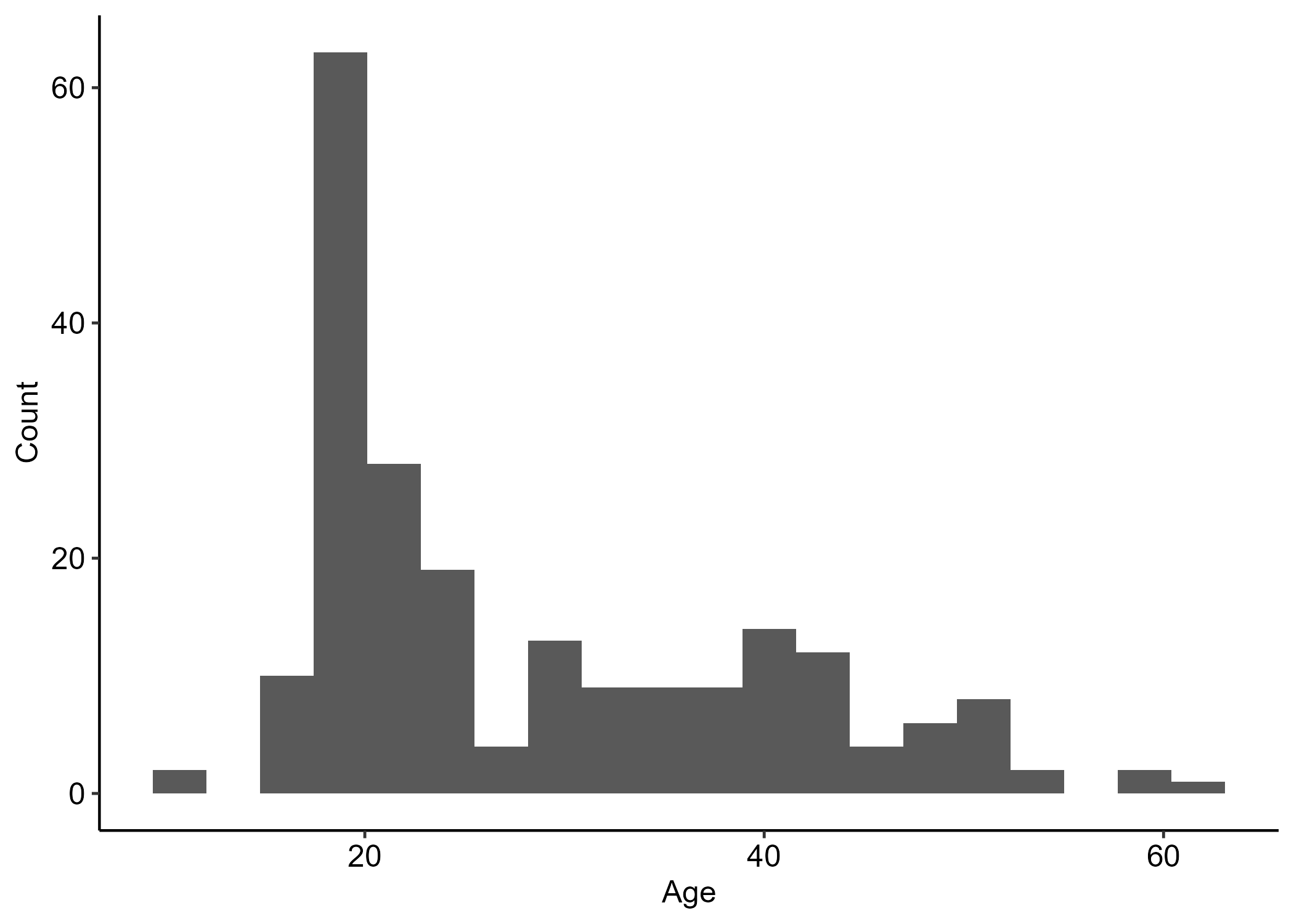}
\caption{Age distribution of participants.}
\label{fig:demographics1}
\end{figure}

Figure~\ref{fig:demographics1} presents the age distribution of the participants. 
The majority of respondents are relatively young, with a concentration in the early twenties. 
The mean age is 28.3 years (SD = 11.3), and the median age is 23 years. 
Participants' ages range from 10 to 61 years (N = 215). 
The distribution is moderately right-skewed due to the presence of several older respondents.

\section{Results}

We start this section by reporting the outcome of the tournaments.

\subsection{Tournaments results}

Table~\ref{tab:top10-points-t1} shows the top-ranked strategies in Tournament 1 with only human participants. The best strategies are clearly non-uniform, but they also avoid extreme concentration on one or two states. Instead, the best-performing allocations spread resources across several states with a handful of pronounced peaks, suggesting that successful play relies on balancing coverage and focus rather than following a naive equal-split rule or an all-in heuristic.

\begin{table}[htbp]
\centering
\caption{Top-10 strategies by points (Tournament 1)}
\label{tab:top10-points-t1}
\resizebox{\textwidth}{!}{%
\begin{tabular}{r l r r r r r r r r r r}
\toprule
Rank & AgentType & Points & State A & State B & State C & State D & State E & State F & State G & State H & State I \\
\midrule
1 & Human & 166.0 & 4 & 13 & 3 & 17 & 21 & 3 & 21 & 5 & 13 \\
2 & Human & 160.0 & 3 & 16 & 3 & 17 & 22 & 17 & 3 & 16 & 3 \\
3 & Human & 156.5 & 2 & 16 & 1 & 17 & 23 & 16 & 2 & 22 & 1 \\
4 & Human & 151.0 & 2 & 12 & 1 & 21 & 20 & 21 & 1 & 20 & 2 \\
5 & Human & 151.0 & 1 & 1 & 1 & 1 & 21 & 21 & 21 & 18 & 15 \\
6 & Human & 150.5 & 13 & 5 & 21 & 13 & 6 & 13 & 4 & 21 & 4 \\
7 & Human & 149.5 & 2 & 1 & 16 & 21 & 2 & 21 & 15 & 21 & 1 \\
8 & Human & 149.0 & 3 & 22 & 21 & 3 & 3 & 21 & 3 & 21 & 3 \\
9 & Human & 148.0 & 12 & 4 & 21 & 3 & 21 & 3 & 3 & 21 & 12 \\
10 & Human & 147.0 & 1 & 18 & 22 & 2 & 2 & 16 & 22 & 15 & 2 \\
\bottomrule
\end{tabular}%
}
\end{table}

When LLM strategies are added in Tournament~2, the strongest performers are still all human: every position in the Top-10 ranking is occupied by a human-submitted strategy (Table~\ref{tab:top10-points-t2}). Despite the second tournament changes the strategic environment by enlarging the opponent pool, yet the set of best responses remains decisively human.

\begin{table}[htbp]
\centering
\caption{Top-10 strategies by points (Tournament 2)}
\label{tab:top10-points-t2}
\resizebox{\textwidth}{!}{%
\begin{tabular}{r l r r r r r r r r r r}
\toprule
Rank & AgentType & Points & State A & State B & State C & State D & State E & State F & State G & State H & State I \\
\midrule
1 & Human & 187.0 & 4 & 13 & 3 & 17 & 21 & 3 & 21 & 5 & 13 \\
2 & Human & 172.5 & 12 & 4 & 21 & 3 & 21 & 3 & 3 & 21 & 12 \\
3 & Human & 172.5 & 2 & 16 & 1 & 17 & 23 & 16 & 2 & 22 & 1 \\
4 & Human & 172.5 & 1 & 23 & 1 & 21 & 21 & 1 & 13 & 16 & 3 \\
5 & Human & 172.0 & 2 & 21 & 2 & 16 & 21 & 1 & 16 & 21 & 0 \\
6 & Human & 169.5 & 16 & 16 & 24 & 3 & 16 & 3 & 3 & 16 & 3 \\
7 & Human & 167.0 & 1 & 18 & 22 & 2 & 2 & 16 & 22 & 15 & 2 \\
8 & Human & 167.0 & 2 & 3 & 12 & 21 & 24 & 21 & 12 & 3 & 2 \\
9 & Human & 166.0 & 2 & 1 & 16 & 21 & 2 & 21 & 15 & 21 & 1 \\
10 & Human & 165.5 & 13 & 5 & 21 & 13 & 6 & 13 & 4 & 21 & 4 \\
\bottomrule
\end{tabular}%
}
\end{table}

 The same pattern persists in Tournament~3, where the number of LLM strategies is matched to the number of human strategies. Table~\ref{tab:top10-points-t3} shows that equalizing the mass of human and LLM strategies does not overturn the main ranking result. Humans again occupy all Top-10 positions. Moreover, several allocations recur across tournaments, including $(4,13,3,17,21,3,21,5,13)$ and $(2,16,1,17,23,16,2,22,1)$, which remain among the strongest strategies in multiple environments. This stability suggests that successful human play is driven more by relatively stable, rule-based heuristics derived from the structure of the game than by strong adaptation to a particular opponent pool.

\begin{table}[htbp]
\centering
\caption{Top-10 strategies by points (Tournament 3)}
\label{tab:top10-points-t3}
\resizebox{\textwidth}{!}{%
\begin{tabular}{r l r r r r r r r r r r}
\toprule
Rank & AgentType & Points & State A & State B & State C & State D & State E & State F & State G & State H & State I \\
\midrule
1 & Human & 357.0 & 2 & 22 & 3 & 14 & 22 & 17 & 15 & 4 & 1 \\
2 & Human & 351.0 & 2 & 16 & 1 & 17 & 23 & 16 & 2 & 22 & 1 \\
3 & Human & 350.0 & 3 & 21 & 1 & 1 & 14 & 19 & 21 & 6 & 14 \\
4 & Human & 350.0 & 4 & 13 & 3 & 17 & 21 & 3 & 21 & 5 & 13 \\
5 & Human & 349.0 & 2 & 1 & 16 & 21 & 2 & 21 & 15 & 21 & 1 \\
6 & Human & 345.0 & 2 & 2 & 17 & 18 & 2 & 21 & 18 & 18 & 2 \\
7 & Human & 343.5 & 2 & 21 & 16 & 21 & 1 & 2 & 0 & 16 & 21 \\
8 & Human & 343.5 & 1 & 18 & 22 & 2 & 2 & 16 & 22 & 15 & 2 \\
9 & Human & 340.0 & 1 & 2 & 20 & 20 & 20 & 20 & 13 & 2 & 2 \\
10 & Human & 340.0 & 21 & 2 & 2 & 2 & 21 & 2 & 14 & 15 & 21 \\
\bottomrule
\end{tabular}%
}
\end{table}

\subsection{Analysis of strategies}


To further characterize the structure of submitted strategies, Table~\ref{tab:t1-reasoning-level} groups allocations by the number of states receiving more than 11 visits. The labels ``Level 0,'' ``Level 1,'' and so on refer to these categories, reported in the first column. The ``Human Percent'' column shows the share of submitted strategies in each category, while the ``Human Score'' column reports the average number of states won per duel by strategies in that category. The table therefore summarizes both how common each allocation pattern is and how well it performs on average.

\begin{table}[htbp]
\centering
\caption{Tournament 1: Distribution by reasoning level}
\label{tab:t1-reasoning-level}
\begin{tabular}{lccc}
\toprule
Reasoning Level & \# of states with $>11$ trips & Human Percent & Human Score \\
\midrule
Level 0 & 1 & 4.4\% & 3.47 \\
Level 1 & 8 & 1.5\% & 4.36 \\
Level 2 & 7 & 5.4\% & 4.47 \\
Level 3 & 6 & 11.2\% & 4.57 \\
Level 4 & 5 & 54.6\% & 4.66 \\
Level 5+ & 4 & 12.2\% & 4.76 \\
Level 5+ & 3 & 7.8\% & 4.13 \\
Level 5+ & 2 & 1.5\% & 2.77 \\
\bottomrule
\end{tabular}
\end{table}

This classification captures one important dimension of strategic sophistication. Following the logic of \citet{arad2012multi}, we treat the number of states with more than 11 visits as a proxy for 
strategy's level of sophistication. Because the budget is 100 and there are nine states, the near-uniform allocation has eight states with 11 trips and one state with 12 trips. We therefore classify allocations with exactly one state receiving more than 11 trips as Level 0. Such strategies do not require a deep analysis of the opponents, and typically they come to mind first. Players who anticipate many Level 0 strategies may look for allocations that beat them. The simplest modification that beats a Level $0$ strategy is sacrificing one state and redistributing, say, one additional visit to any other states. We therefore classify strategies with eight states receiving more than 11 trips as Level 1. By iterating this way of thinking, we reach Level $k$ strategies, $k\geqslant 2$, each additional level corresponds to sacrificing one more state and reallocating trips to the remaining states. In the human-only tournament 1, the majority of strategies were Level $4$ strategies, accounting for 54.6\% of all human strategies (Table~\ref{tab:t1-reasoning-level}). Strategies with only one state above the threshold are rare and perform relatively poorly, with an average score of 3.47 compared with 4.66 for the Level $4$ category and 4.76 for Level $5$ category. At the same time, the highest average score in Tournament~1 is achieved by Level $5$ strategies, following by Level $4$ and Level $3$ strategies. It suggests that empirically the most successful human play involves the intermediate level of concentration. Note that one should carefully interpret strategies with high values of $k$: starting from $k=5$, the logic of sacrificing one additional state without taking into account what happens in the sacrificed states, leads to a loss. Therefore, we introduce a separate category ``level 5+'' strategies --- the strategies with no more than 4 strong states.  

\begin{table}[htbp]
\centering
\caption{Tournament 1: Unit-digit distribution}
\label{tab:t1-table2-like}
\begin{tabular}{lcccccccccc}
\toprule
 & \multicolumn{10}{c}{Unit digit} \\
\cmidrule(lr){2-11}
Agent & 0 & 1 & 2 & 3 & 4 & 5 & 6 & 7 & 8 & 9 \\
\midrule
Human & 28.6\% & 19.0\% & 14.3\% & 8.8\% & 5.4\% & 6.9\% & 5.6\% & 2.8\% & 4.6\% & 4.1\% \\
\bottomrule
\end{tabular}
\end{table}

Table~\ref{tab:t1-table2-like} reports the distribution of unit digits in state-level allocations. For each digit from 0 to 9, the table shows the share of all state-level allocations ending in that digit. In Tournament~1, the distribution already points to a human tendency to rely on salient numbers, especially digits 0, 1, and 2. However, this tendency is moderate rather than overwhelming: human allocations still use the other digits with non-trivial frequency, which is consistent with flexible but imperfect heuristic reasoning rather than rigid templating.

\begin{figure}[htbp]
    \centering
    \includegraphics[width=0.8\linewidth]{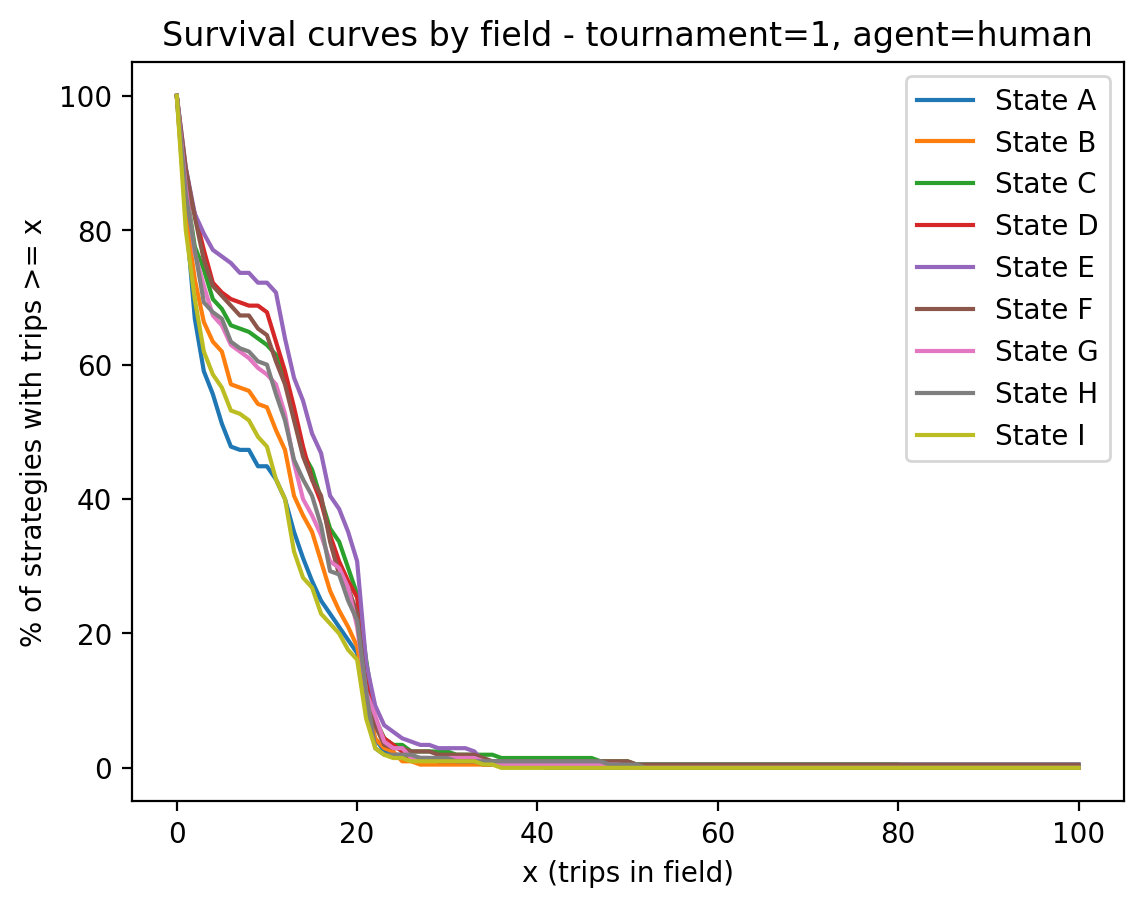}
\caption{Survival rate: Human (Tournament 1)}
\label{fig:survival1_human}
\end{figure}

Figure~\ref{fig:survival1_human} complements this interpretation. The survival curves for human allocations are relatively smooth and similar across states, indicating that players typically choose from a common family of strategies. Human behavior is therefore dispersed enough to avoid collapsing on a single focal allocation, but structured enough to generate a recognizable distribution of play.

\begin{table}[htbp]
\centering
\caption{Tournament 2: Distribution by reasoning level}
\label{tab:t2-reasoning-level}
\begin{tabular}{lccccc}
\toprule
Reasoning Level & \# of states with $>11$ trips & \multicolumn{2}{c}{Human} & \multicolumn{2}{c}{LLM} \\
\cmidrule(lr){3-4}\cmidrule(lr){5-6}
 &  & Percent & Score & Percent & Score \\
\midrule
Level 0 & 1 & 5.8\% & 3.18 & 17.4\% & 4.40 \\
Level 1 & 8 & 1.4\% & 4.63 & 4.3\% & 4.70 \\
Level 2 & 7 & 3.4\% & 4.72 & 8.7\% & 4.67 \\
Level 3 & 6 & 11.1\% & 4.72 & 0.0\% & -- \\
Level 4 & 5 & 50.2\% & 4.71 & 17.4\% & 4.46 \\
Level 5+ & 4 & 16.9\% & 4.69 & 8.7\% & 4.32 \\
Level 5+ & 3 & 5.8\% & 4.01 & 13.0\% & 4.45 \\
Level 5+ & 2 & 4.3\% & 3.68 & 4.3\% & 4.08 \\
\bottomrule
\end{tabular}
\end{table}

Tournament~2 reveals a clear difference between humans and LLMs. Human strategies remain concentrated in the middle of the reasoning distribution, with 50.2\% of strategies placing more than 11 trips in exactly five states (Table~\ref{tab:t2-reasoning-level}). LLM strategies are much less centered: 17.4\% fall into the lowest category, another 17.4\% into the modal human category, and none appear in the six-state category. The performance numbers also favor humans overall. Averaging across all submitted strategies, humans win about 4.52 states per duel, compared with roughly 4.36 for LLMs. The gap is not enormous, but it is systematic and large enough to keep LLMs out of the top of the standings.

\begin{table}[htbp]
\centering
\caption{Tournament 2: Unit-digit distribution}
\label{tab:t2-table2-like}
\begin{tabular}{lcccccccccc}
\toprule
 & \multicolumn{10}{c}{Unit digit} \\
\cmidrule(lr){2-11}
Agent & 0 & 1 & 2 & 3 & 4 & 5 & 6 & 7 & 8 & 9 \\
\midrule
Human & 30.0\% & 18.3\% & 12.3\% & 8.8\% & 5.5\% & 7.9\% & 5.9\% & 4.1\% & 3.7\% & 3.6\% \\
LLM & 30.4\% & 36.7\% & 21.7\% & 3.4\% & 2.4\% & 0.5\% & 1.0\% & 1.4\% & 1.4\% & 1.0\% \\
\bottomrule
\end{tabular}
\end{table}

The unit-digit evidence also highlights a clear difference between humans and LLMs. In Tournament~2, 88.8\% of LLM state-level allocations end in 0, 1, or 2, and digit 1 alone accounts for 36.7\% of all LLM choices (Table~\ref{tab:t2-table2-like}). Human allocations also display focality, but they are less concentrated, with substantially more mass on digits 3 through 9. This pattern suggests that LLMs rely more heavily on a narrow set of salient numerical choices. One possible reason is the attractiveness of allocations such as 11, which provide a natural benchmark for splitting 100 trips across nine states.

\begin{figure}[htbp]
    \centering
    \includegraphics[width=0.48\linewidth]{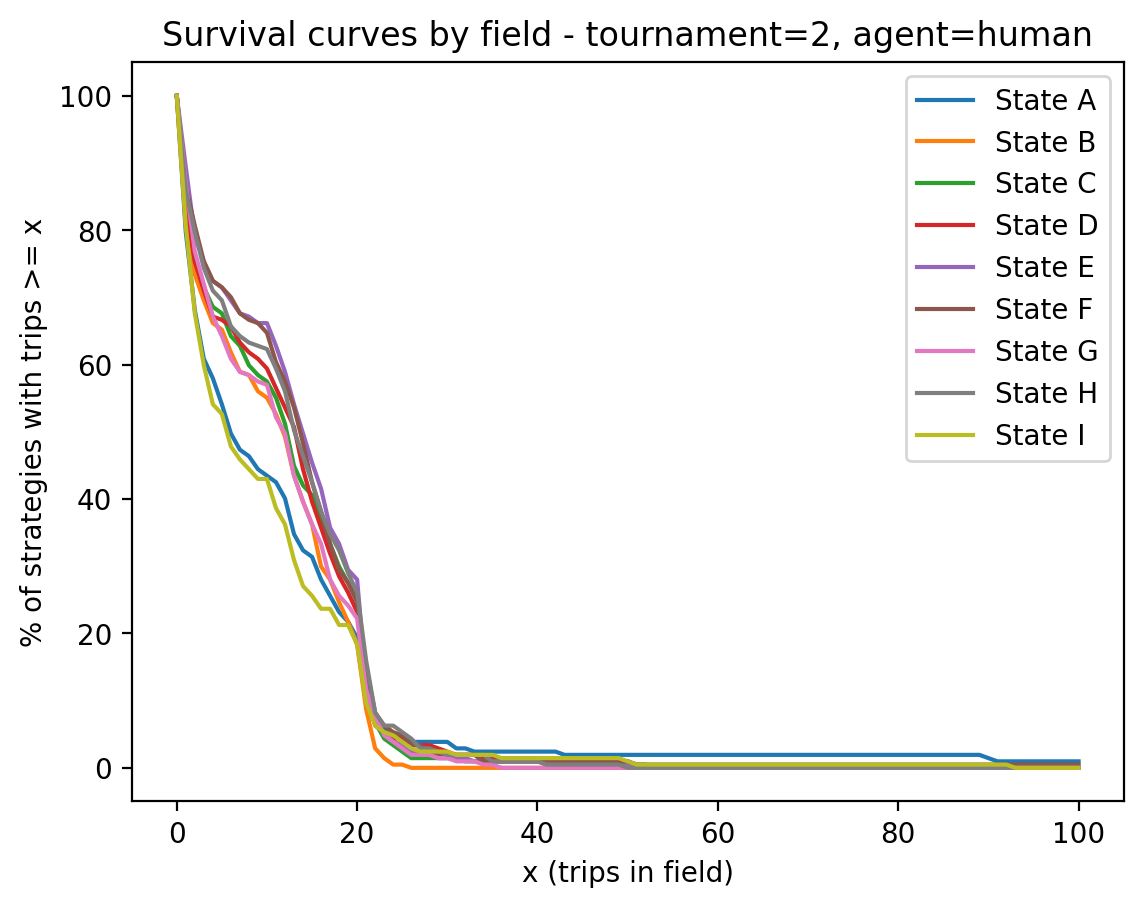}
    \hfill
    \includegraphics[width=0.48\linewidth]{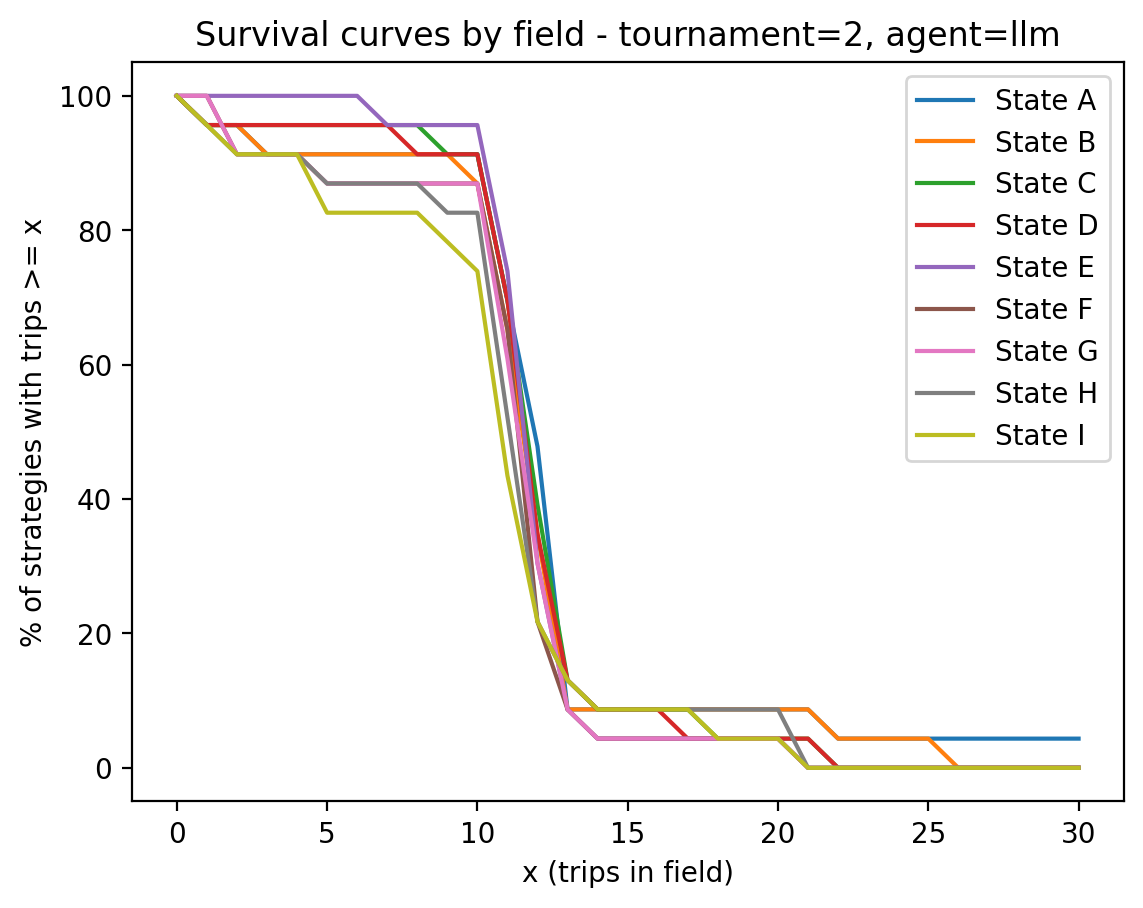}
    \caption{Survival rate: Human vs LLM (Tournament 2)}
    \label{fig:survival2_joint}
\end{figure}

The survival curves in Figure~\ref{fig:survival2_joint} visualize the same contrast in a different way. Human curves decline gradually and exhibit a broad range of allocations across states. LLM curves, by contrast, show a pronounced cliff around 10 to 12 trips, indicating that many model-generated strategies bunch around similar cutoffs. This kind of bunching is precisely what one would expect from heuristic generation based on salient numbers rather than from richer opponent-contingent reasoning.

\begin{table}[htbp]
\centering
\caption{Tournament 3: Distribution by reasoning level}
\label{tab:t3-reasoning-level}
\begin{tabular}{lccccc}
\toprule
Reasoning Level & \# of states with $>11$ trips & \multicolumn{2}{c}{Human} & \multicolumn{2}{c}{LLM} \\
\cmidrule(lr){3-4}\cmidrule(lr){5-6}
 &  & Percent & Score & Percent & Score \\
\midrule
Level 0 & 1 & 6.8\% & 3.29 & 27.1\% & 4.14 \\
Level 1 & 8 & 1.4\% & 5.62 & 5.8\% & 5.53 \\
Level 2 & 7 & 5.8\% & 5.58 & 3.9\% & 5.30 \\
Level 3 & 6 & 9.2\% & 5.20 & 3.9\% & 5.21 \\
Level 4 & 5 & 48.3\% & 4.86 & 18.8\% & 4.74 \\
Level 5+ & 4 & 16.4\% & 4.54 & 4.8\% & 4.60 \\
Level 5+ & 3 & 8.7\% & 3.98 & 2.9\% & 4.40 \\
Level 5+ & 2 & 1.4\% & 2.82 & 5.3\% & 4.13 \\
\bottomrule
\end{tabular}
\end{table}

Tournament~3 strengthens these conclusions because the number of LLM strategies is now the same as the number of human strategies. If the earlier results were driven mainly by composition effects, the human advantage should weaken substantially here. It does not. Humans again outperform LLMs on average, winning about 4.63 states per duel against 4.37 for LLMs. As evident from the Table~\ref{tab:t3-reasoning-level}, the distributional contrast also becomes starker: 27.1\% of LLM strategies fall into the lowest reasoning category, compared with only 6.8\% of human strategies, while humans remain heavily concentrated in the five-state category (48.3\%). Thus, even after controlling for the relative mass of model-generated strategies in the tournament population, human play remains both more effective and more regularly centered around the middle-to-high reasoning levels.

\begin{table}[htbp]
\centering
\caption{Tournament 3: Unit-digit distribution}
\label{tab:t3-table2-like}
\begin{tabular}{lcccccccccc}
\toprule
 & \multicolumn{10}{c}{Unit digit} \\
\cmidrule(lr){2-11}
Agent & 0 & 1 & 2 & 3 & 4 & 5 & 6 & 7 & 8 & 9 \\
\midrule
Human & 30.3\% & 18.5\% & 12.6\% & 9.4\% & 5.7\% & 7.6\% & 6.8\% & 3.3\% & 3.3\% & 2.7\% \\
LLM & 23.9\% & 42.9\% & 18.0\% & 2.0\% & 2.9\% & 2.7\% & 1.8\% & 1.1\% & 3.4\% & 1.2\% \\
\bottomrule
\end{tabular}
\end{table}

The digit patterns in Tournament~3 are even more striking. Among LLMs, 84.8\% of all state-level allocations end in 0, 1, or 2, and the digit 1 alone appears in 42.9\% of cases (Table~\ref{tab:t3-table2-like}). Human strategies again use these digits frequently, but their distribution is far less concentrated. This reinforces the interpretation that LLMs can produce coherent-looking Blotto plans at a coarse level while still relying on overly stereotyped numerical choices at the fine-grained level.

\begin{figure}[htbp]
    \centering
    \includegraphics[width=0.48\linewidth]{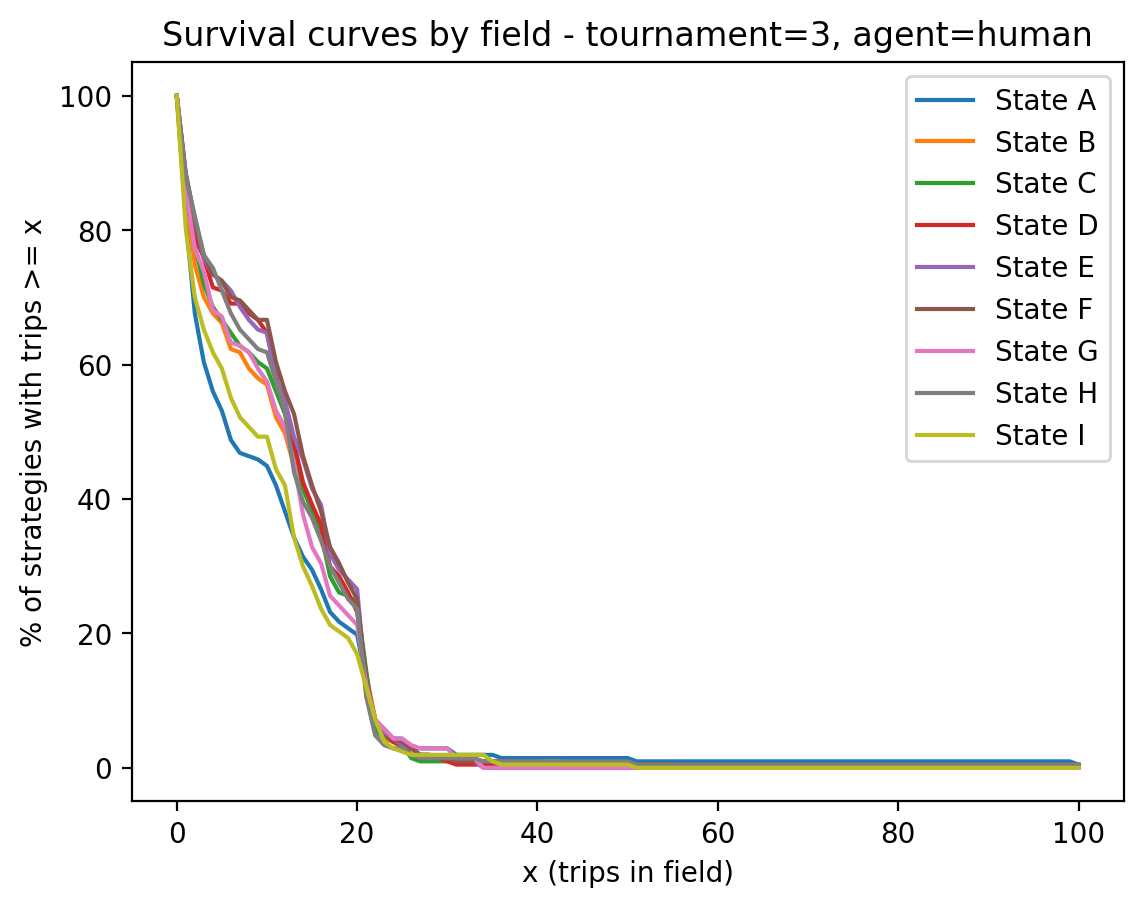}
    \hfill
    \includegraphics[width=0.48\linewidth]{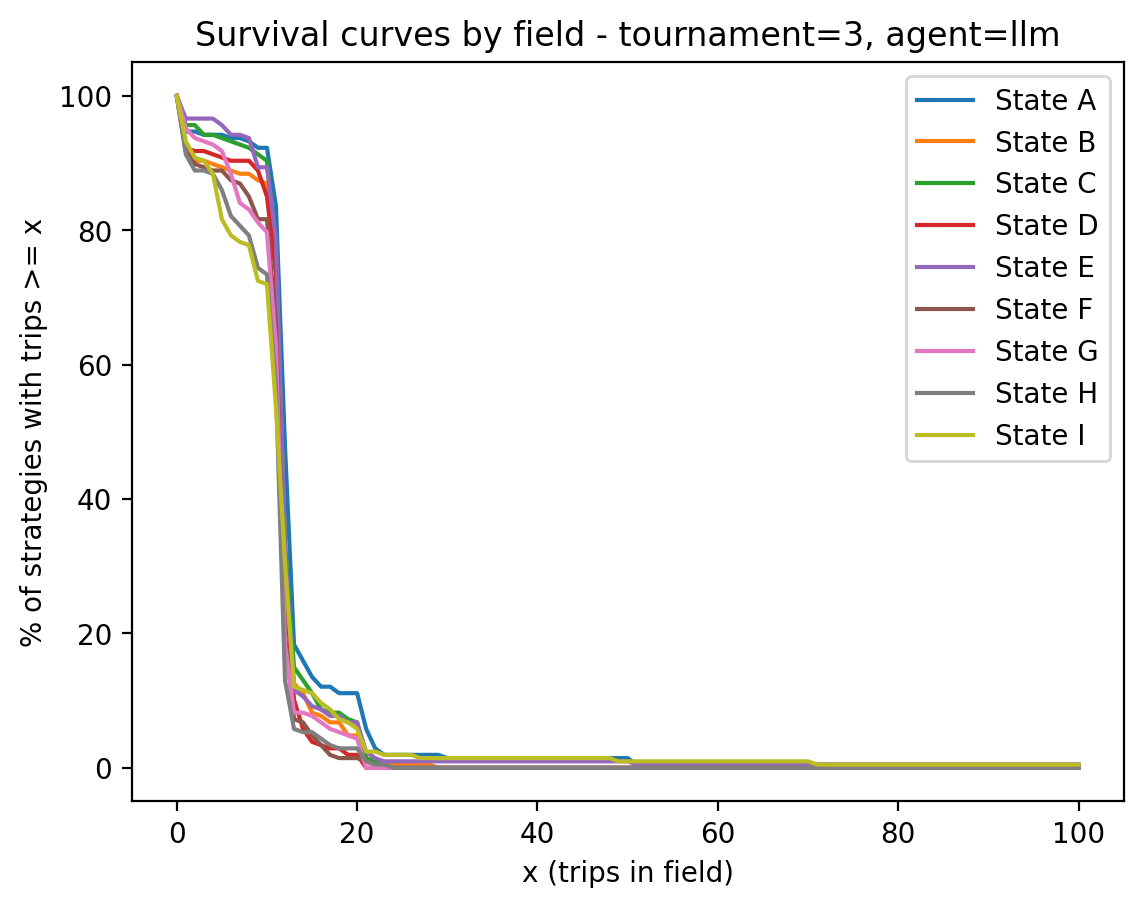}
    \caption{Survival rate: Human vs LLM (Tournament 3)}
    \label{fig:survival3_joint}
\end{figure}

Figure~\ref{fig:survival3_joint} confirms that this difference survives in the balanced tournament. Human allocations remain heterogeneous but smooth, whereas LLM strategies continue to cluster tightly around a narrow set of values. Taken together, the ranking tables, reasoning-level distributions, digit frequencies, and survival curves all point in the same direction: humans do not merely outperform LLMs by chance or by tournament composition. They appear to use more effective and less stereotyped heuristics in this high-dimensional strategic environment.

\subsection{LLM leader board}

Table~\ref{tab:llm-rating-t3} reports the ranking of LLM families in Tournament~3 based on their average tournament points across all submitted instances. The table shows substantial heterogeneity in performance across models. The strongest average results are achieved by \texttt{google/gemini-2.5-pro}, \texttt{x-ai/grok-4}, and \texttt{openai/gpt-5}, all of which clearly outperform the rest of the model pool. A second tier includes models such as \texttt{openai/o3}, \texttt{qwen/qwen3-max}, and \texttt{moonshotai/kimi-k2-thinking}, while smaller and older open-weight models tend to occupy the lower part of the ranking.

At the same time, the number of instances differs across models, and this difference should be taken into account when interpreting the leaderboard. In Tournament~3, we aimed to construct a balanced pool of 207 LLM strategies, matching the number of valid human strategies, and to give each model a comparable representation in the competition. In particular, the initial design targeted approximately eight instances per model. 

However, not all model outputs could be retained. Some generated allocations violated the rules of the game, for example, by failing to produce a valid integer allocation across the nine states or by exceeding the total budget constraint. Such invalid submissions were removed from the dataset. To restore the target pool size of 207 strategies, replacement submissions were then generated using randomly selected models. As a result, some models ended up contributing more valid instances than initially planned, while others contributed fewer.

Therefore, the \textit{Instances} column should not be interpreted as a deliberate measure of model importance or sampling weight. Instead, it reflects the combination of two factors: first, whether a model's originally generated strategies satisfied the rules of the game; and second, whether that model happened to be randomly selected to supply replacement strategies after invalid submissions had been removed. At the same time, the table suggests that failure to comply with the rules was itself informative about model quality: models that produced more invalid submissions also tended to appear lower in the rating. This pattern indicates that weaker-performing models were not only less successful strategically, but were also less reliable in producing admissible strategies.

\begin{table}[htbp]
\centering
\caption{LLM leader board (Tournament 3)}
\label{tab:llm-rating-t3}
\begin{tabular}{rlrr}
\toprule
Position & Model & Avg. Points & Instances \\
\midrule
1 & google/gemini-2.5-pro & 294.056 & 9 \\
2 & x-ai/grok-4 & 290.5 & 9 \\
3 & openai/gpt-5 & 276.85 & 10 \\
4 & openai/o3 & 267.5 & 8 \\
5 & qwen/qwen3-max & 237.812 & 8 \\
6 & moonshotai/kimi-k2-thinking & 210.444 & 9 \\
7 & deepseek/deepseek-v3.2-exp & 209.111 & 9 \\
8 & mistralai/mistral-medium-3.1 & 191 & 8 \\
9 & z-ai/glm-4.6 & 190.938 & 8 \\
10 & qwen/qwen3-235b-a22b-thinking-2507 & 183.75 & 8 \\
11 & meta-llama/llama-3.2-3b-instruct:free & 182 & 1 \\
12 & anthropic/claude-sonnet-4.5 & 178 & 12 \\
13 & openai/gpt-oss-120b & 173.5 & 8 \\
14 & qwen/qwen3-8b & 169.15 & 10 \\
15 & meta-llama/llama-3.1-405b-instruct & 155.722 & 9 \\
16 & minimax/minimax-m2 & 147.167 & 9 \\
17 & qwen/qwen3-14b & 147.167 & 6 \\
18 & mistralai/mistral-small-3.1-24b-instruct & 137.111 & 9 \\
19 & meta-llama/llama-3.3-70b-instruct & 132 & 9 \\
20 & moonshotai/kimi-k2-0905 & 129.188 & 8 \\
21 & qwen/qwen3-next-80b-a3b-thinking & 110.929 & 7 \\
22 & openai/gpt-oss-20b & 92.75 & 6 \\
23 & mistralai/mistral-7b-instruct & 91.667 & 9 \\
24 & qwen/qwen3-32b & 78.625 & 8 \\
25 & meta-llama/llama-3.2-1b-instruct & 71.5 & 5 \\
26 & meta-llama/llama-3.1-8b-instruct & 67.3 & 5 \\
\bottomrule
\end{tabular}
\end{table}

\FloatBarrier
\section{Modeling the Relationship Between Reasoning and Performance}

\subsection{Regression equation}

To assess how the payoff consequences of reasoning levels vary across tournaments, we estimate the following regression model:

\begin{equation}
Y_{ir} = \alpha + \sum_{k=1}^{5} \beta_k \, ReasoningLevel_{ikr} + 
\gamma_1 Age_i + \gamma_2 Age_i^2 + \delta Female_i + 
\theta X_i + \varepsilon_{ir}.
\end{equation}

Here $Y_{ir}$ denotes the number of points obtained by participant $i$ in round $r$ of the tournament ($r \in \{1,2,3\}$). 
$ReasoningLevel_{ikr}$ is a set of indicator variables capturing the participant’s reasoning level, where Level~0 serves as the reference category. 
$X_i$ represents a vector of additional controls, including education, field of study, and employment status. 
Robust standard errors are reported in parentheses.

\subsection{Regression results}

Table~\ref{tab:reasoning_rounds} reports the relationship between reasoning level and performance across three tournaments. The dependent variables are the points obtained in the first, second, and third tournaments, respectively. The omitted category for the reasoning variable is Level~0.

\begin{table}[!htbp]\centering
\caption{Reasoning Level and Performance Across Tournament Rounds}
\label{tab:reasoning_rounds}
\resizebox{\textwidth}{!}{
\begin{tabular}{lccc}
\toprule
 & \multicolumn{3}{c}{\textit{Dependent variable}} \\
\cmidrule(lr){2-4}
 & First Round Points & Second Round Points & Third Round Points \\
 & (1) & (2) & (3) \\
\midrule

\textit{Reasoning Level (ref: Level 0)} \\

Level 1 & $-$1.09 & $-$2.61 & $-$1.85 \\
 & (23.77) & (22.77) & (20.30) \\

Level 2 & $-$1.82 & 1.15 & $-$0.36 \\
 & (15.40) & (16.59) & (12.35) \\

Level 3 & 4.41 & 8.82 & 8.26 \\
 & (13.15) & (12.07) & (10.95) \\

Level 4 & 34.17$^{***}$ & 41.16$^{***}$ & 43.88$^{***}$ \\
 & (11.26) & (10.20) & (8.68) \\

Level 5+ & 23.35$^{*}$ & 13.77 & $-$0.77 \\
 & (11.96) & (10.60) & (8.98) \\

\addlinespace
Age & 3.17 & 2.01 & 0.01 \\
 & (2.03) & (1.91) & (1.72) \\

Age$^2$ & $-$0.05$^{*}$ & $-$0.03 & $-$0.003 \\
 & (0.03) & (0.03) & (0.02) \\

Female & $-$7.31 & $-$4.70 & $-$0.53 \\
 & (6.54) & (6.27) & (5.84) \\

\addlinespace
\textit{Education (ref: Higher Education)} \\

Doctoral degree & $-$0.12 & $-$2.20 & 0.84 \\
 & (10.49) & (10.27) & (9.45) \\

Secondary Education & 2.19 & 7.28 & 4.94 \\
 & (9.91) & (9.35) & (8.51) \\

\addlinespace
\textit{Field of Study (ref: Humanities, Social Sciences, and other)} \\

Economics and Management & 6.24 & $-$2.76 & $-$6.59 \\
 & (9.12) & (8.74) & (8.07) \\

STEM & 14.01$^{*}$ & 6.37 & 0.68 \\
 & (8.40) & (8.06) & (7.51) \\

\addlinespace
\textit{Employment Status (ref: Not Working)} \\

Student & 0.46 & 2.95 & $-$3.10 \\
 & (12.46) & (12.06) & (11.14) \\

Employed & $-$0.27 & 1.67 & 4.85 \\
 & (9.78) & (9.48) & (8.59) \\

\addlinespace
Constant & $-$51.52 & $-$30.64 & 4.99 \\
 & (39.43) & (36.50) & (32.33) \\

\midrule
Observations & 205 & 207 & 207 \\
$R^{2}$ & 0.19 & 0.24 & 0.36 \\
Adjusted $R^{2}$ & 0.13 & 0.19 & 0.31 \\
Residual Std. Error & 35.93 & 34.69 & 31.82 \\
F Statistic & 3.24$^{***}$ & 4.43$^{***}$ & 7.58$^{***}$ \\

\bottomrule
\multicolumn{4}{l}{\footnotesize Robust standard errors in parentheses.} \\
\multicolumn{4}{l}{\footnotesize $^{*}p<0.1$, $^{**}p<0.05$, $^{***}p<0.01$.}
\end{tabular}
}
\end{table}

The results indicate a strong and robust positive association between Level 4 reasoning and performance. These strategies substantially outperform Level~0 across all three tournaments. The estimated coefficients range from 34 to 44 additional points in the tournament and are statistically significant at the 1\% level in all specifications. 

Though strategies classified as Level~5+ also exhibit a performance advantage in the first tournament, the estimates are smaller, the significance level is lower, and they become statistically insignificant in tournaments 2 and 3.

Lower reasoning levels (Levels~1--3) do not differ significantly from the baseline. Their coefficients are small and statistically insignificant across all rounds, indicating that moderate increases in reasoning depth do not systematically translate into improved performance. 

We conclude that it is the specific property of Level 4 strategies --- having 5 strong and 4 weak but non-zero states --- that leads to success in our tournaments. In general, it is not true that strategies with higher level of reasoning are better.

The control variables have limited explanatory power. Age displays a weak concave relationship with performance in the first tournament, as reflected by the negative coefficient on age squared, but this pattern does not persist in tournaments 2 and 3. Gender, education, and employment status are not significantly associated with outcomes. The only notable pattern outside reasoning levels is that participants with STEM backgrounds perform better in the first tournament.

Overall, the findings suggest that strategic sophistication is key to success if and only if the necessary level of reasoning depth is reached. While lower levels of reasoning offer no clear advantage over the baseline, Level 4 is associated with substantially improved performance across all rounds. We emphasize that a Level 4 strategy requires non-zero numbers in weak states, thus differing significantly from more trivial strategies that allocate all visits to five states.

\section{Discussion}

This paper studies strategic behavior in a Colonel Blotto environment. The experimental design allows us to compare the performance of different classes of strategies, including strategies of different reasoning levels and strategies submitted by human participants versus those generated by large language models. We find that best-performing strategies combine selective concentration on the majority of states with positive allocations to weaker states.

This pattern is closely related to the logic of iterative reasoning. In a Blotto contest with nine symmetric states and a budget of 100, the equal-split allocation provides a natural Level 0 benchmark. A player who expects many opponents to use such a benchmark may profitably sacrifice one state and allocate slightly more than the equal split to the remaining states. Further iterations of this logic generate strategies with fewer strong states and higher allocations to those states. Our empirical results show that this reasoning is useful only up to a point. Level 4 strategies, which typically allocate more than 11 trips to five states while keeping non-zero allocations elsewhere, consistently perform well. Lower or higher levels do not generate a statistically significant advantage over the baseline. The resulting relationship between reasoning depth and performance is therefore non-monotonic. 

The comparison with LLM-generated strategies reinforces this interpretation. LLMs frequently produce coherent-looking allocations, but these allocations are more stereotyped than human ones. In particular, LLM strategies display stronger bunching around salient numerical values and are less consistently represented in the empirically successful reasoning categories. LLM play is structured in a relatively narrow and focal way. This makes LLMs an informative benchmark. Their weaker performance does not appear to reflect random or incoherent play among valid submissions; rather, it is consistent with reliance on allocation templates that are insufficiently adapted to the strategic nature of the tournament.

This distinction is important for interpreting the human advantage. Humans do not appear to display strong opponent-specific adaptation: despite facing different opponent pools across tournaments, participants changed their strategies only modestly. Their advantage instead seems to come from a class of robust heuristics that are well suited to this particular empirical environment. These heuristics are structured enough to exploit naive allocations, but heterogeneous enough to avoid the most obvious forms of predictability.

The paper therefore contributes to the experimental literature on Colonel Blotto games by introducing LLMs as a new comparison group in a complex strategic environment. Because LLMs generate strategies that are systematically different from human strategies, they help reveal which features of human play lead to competitive advantage.

The results also speak to the broader use of LLMs as simulated economic agents. In this experiment, LLMs are not close substitutes for human participants: they do not match the dispersion, calibration, and robustness of human strategies. This suggests caution in using LLMs as stand-ins for human subjects in arbitrary strategic settings, especially when the relevant behavior depends on higher-dimensional reasoning and on beliefs about heterogeneous opponents.

Several limitations should be emphasized. The reasoning-level classification is based on observable properties of strategies rather than on direct observation of cognitive processes. Different participants may arrive at the same allocation through different reasoning paths. Similarly, LLM performance depends on the specific models, prompts, and sampling procedure used in the experiment. Alternative prompts, repeated sampling, or feedback could change model behavior. 

Overall, the evidence suggests that in a high-dimensional Blotto tournament, successful play is driven by choosing a well-calibrated reasoning depth rather than maximizing reasoning depth. Humans outperform LLMs because their heuristics are better calibrated to the distribution of opponents’ strategies. The central lesson is not simply that humans beat current LLMs, but that the structure of human heuristics matters: intermediate concentration, positive allocation to weak states, and avoidance of overly stereotyped numerical patterns are all competitively valuable.

\section{Conclusion}

We study behavior in a discrete Colonel Blotto tournament with human participants and large language models. The main result is that humans outperform LLM-generated strategies. This result holds both when several LLM strategies are added to the tournament, and when the number of LLM strategies is matched to the number of human strategies. The human advantage is therefore not a mechanical consequence of tournament composition.

The analysis suggests that performance is driven by the structure of submitted strategies. The most successful human allocations use an intermediate degree of concentration: they create 5 strong states while maintaining positive allocations to 4 weaker states. Regression results show that such Level 4 strategies are robustly associated with better performance across tournaments. Thus, strategic reasoning matters, but its payoff effect is non-monotonic.

LLM strategies differ from human strategies in ways that are consequential for performance. They are more concentrated around salient numerical templates and less consistently located in the empirically successful region of the strategy space. The models often generate admissible and plausible strategies, but these strategies are less well calibrated to the distribution of opponents’ play. In this sense, the experiment identifies a specific limitation of current LLMs in strategic environments: they can imitate the form of strategic reasoning without reliably producing the allocation patterns that perform well in a complex tournament.

The findings contribute to the emerging literature on LLMs as economic agents by showing that human and LLM behavior can diverge substantially in high-dimensional strategic games. In our setting, LLMs are useful as a comparison group, but they are not reliable substitutes for human subjects.

\section*{Funding}

The study was supported by a grant from the Russian Science Foundation, No. 25-18-00539, \url{https://rscf.ru/project/25-18-00539/}.

\section*{Declaration of generative AI and AI-assisted technologies in the writing process}

During the preparation of this work, the authors used LLM models as economic agents in their experiments. GPT-5 was used for language proofreading purposes. After using these tools, the authors reviewed and edited the content as needed and take full responsibility for the content of the published article.

\section*{Data availability}

The data supporting the findings of this study is publicly available on Harvard Dataverse: \url{https://doi.org/10.7910/DVN/YUM1BI}.

\bibliographystyle{apa-good}
\bibliography{sample-bibliography}

\section*{Appendix}

\appendix
\appendix
\section{Survey instrument (English translation)}

\subsection*{Page 1}

We invite you to take part in the tournament for the game ``Pre-Election Race''!

There will be a total of 3 tournaments, each of which differs from the others only in the set of participants.

In the first tournament, strategies created by humans will compete.

In the second tournament, to the strategies submitted by humans we will add one strategy from several popular large language models.

In the third tournament, large language models will submit, in total, as many strategies as humans create.

You may submit your strategy to each of the three tournaments; these three strategies may be different.

In each tournament, every submitted strategy will play the game ``Pre-Election Race'' once against every other strategy.

A win in an individual match gives 1 point, a draw gives 0.5 points, and a loss gives 0 points.

In each tournament, final results are determined by the number of points earned in all matches of that tournament.

Based on the results of the three tournaments, overall standings will be determined.

The top 100 places in each tournament will receive prize points:

\begin{itemize}
\item 1st place --- 200 points
\item 2nd place --- 180 points
\item 3rd place --- 165 points
\item 4th place --- 150 points
\item 5th place --- 140 points
\item 6th place --- 130 points
\item 7th place --- 120 points
\item 8th place --- 110 points
\item 9th place --- 100 points
\item Place X from 10 to 100 --- $101 - X$ points
\end{itemize}

In case of ties, the participants share equally the points for the corresponding places.

In Tournaments 2 and 3, ranking is calculated taking into account strategies submitted by language models.

In the overall standings, the winners are those who have earned the highest number of prize points across all three tournaments!

The top 30 participants will receive a gift --- popular science literature with an autograph of <ARTICLE\_AUTHOR\_NAME> (delivery within Russia).

The participant who takes first place will be able to meet with the tournament organizers over lunch in Moscow or, if schedules coincide, in other cities in Russia.

If an in-person meeting is not possible, an online meeting can be arranged.

Good luck!

\subsection*{Page 2}

\begin{enumerate}
\item Two candidates compete for the presidency of a fictional overseas country.
\item The overseas country has 9 states: A, B, C, D, E, F, G, H, I.
\item Each candidate has resources for 100 campaign trips.
\item Each of the two candidates simultaneously and independently decides how many times and to which states to travel. Each state can be visited any integer number of times from 0 to 100.
\item In each state, the candidate who visited that state more times wins. For winning a state, a candidate receives 1 electoral vote. If both candidates visited a state the same number of times, the result in that state is a draw, and both receive 0.5 votes.
\item The president is the candidate who receives more electoral votes. If both candidates receive the same number of votes, they toss a fair coin at the Central Election Commission, i.e., each becomes president with probability 0.5.
\end{enumerate}

\paragraph{Tournament 1}

Please indicate, for each of the nine states A, B, C, D, E, F, G, H, I, how many trips you will make to that state.

In total, you may make \textbf{no more than 100 trips}.

In this tournament, only strategies submitted by other people will play against you.

\begin{center}
\begin{tabular}{lllll}
State A: \underline{\hspace{2cm}} & State B: \underline{\hspace{2cm}} & State C: \underline{\hspace{2cm}} \\
State D: \underline{\hspace{2cm}} & State E: \underline{\hspace{2cm}} & State F: \underline{\hspace{2cm}} \\
State G: \underline{\hspace{2cm}} & State H: \underline{\hspace{2cm}} & State I: \underline{\hspace{2cm}} \\
\end{tabular}
\end{center}

\subsection*{Page 3}

\paragraph{Tournament 2}

Please indicate, for each of the nine states A, B, C, D, E, F, G, H, I, how many trips you will make to that state.

In total, you may make \textbf{no more than 100 trips}.

In this tournament, you will play against both strategies submitted by other people and one strategy from several popular modern large language models.

\begin{center}
\begin{tabular}{lllll}
State A: \underline{\hspace{2cm}} & State B: \underline{\hspace{2cm}} & State C: \underline{\hspace{2cm}} \\
State D: \underline{\hspace{2cm}} & State E: \underline{\hspace{2cm}} & State F: \underline{\hspace{2cm}} \\
State G: \underline{\hspace{2cm}} & State H: \underline{\hspace{2cm}} & State I: \underline{\hspace{2cm}} \\
\end{tabular}
\end{center}

\subsection*{Page 4}

\paragraph{Attention Check Question}

From the dropdown menu, select \textbf{``Yellow''}.

\textit{Response options:} Blue; Black; Orange; Purple; Yellow; Pink.

\paragraph{Tournament 3}

Please indicate, for each of the nine states A, B, C, D, E, F, G, H, I, how many trips you will make to that state.

In total, you may make \textbf{no more than 100 trips}.

In this tournament, you will play against both strategies submitted by other people and strategies from several popular modern large language models.

At the same time, the number of strategies from large language models will be equal to the number of strategies submitted by humans.

\begin{center}
\begin{tabular}{lllll}
State A: \underline{\hspace{2cm}} & State B: \underline{\hspace{2cm}} & State C: \underline{\hspace{2cm}} \\
State D: \underline{\hspace{2cm}} & State E: \underline{\hspace{2cm}} & State F: \underline{\hspace{2cm}} \\
State G: \underline{\hspace{2cm}} & State H: \underline{\hspace{2cm}} & State I: \underline{\hspace{2cm}} \\
\end{tabular}
\end{center}

\subsection*{Page 5}

\paragraph{Age}
Please indicate your age.\\
\textit{Response format:} integer.

\paragraph{Gender}
Please indicate your gender.\\
\textit{Response options:} Male; Female.

\paragraph{Education}
Please indicate your level of education.\\
\textit{Response options:} Incomplete secondary; Secondary; Incomplete higher; Higher; Doctoral degree.

\paragraph{Current occupation}
How would you describe your current occupation?\\
\textit{Response options:}
Studying (school student, university student, PhD student);
Employed (in an organization);
Self-employed / freelancer;
Entrepreneur / business owner;
Temporarily not working (job search, parental leave, etc.);
Retired;
Other.

\paragraph{Field closest to you}
Which field of knowledge is closest to you?\\
\textit{Response options:}
Mathematics;
Computer science and technical sciences;
Economics, business and management;
Natural sciences (physics, chemistry, biology, etc.);
Social sciences (sociology, psychology, political science, etc.).

\paragraph{Decision-making principle}
Describe your decision-making principle.\\
\textit{Explanation:}
What did you rely on when choosing a strategy? Did you change your strategy depending on the set of opponents? Describe your logic.

\paragraph{Email}
Please provide your email address.

\section{LLM prompts (English translation)}

\subsection{Tournament 2}

\paragraph{System message}
\begin{quote}\ttfamily
You are a tournament participant. Reply ONLY with a JSON object strictly following this schema:

\{"A": <int>, "B": <int>, "C": <int>, "D": <int>, "E": <int>, "F": <int>, "G": <int>, "H": <int>, "I": <int>, "total": <int $\leq$ 100>, "explanation": <string in Russian>\}

The total number of trips must be $\leq 100$. Do not add any text, any symbols outside the JSON object, or any Markdown.
\end{quote}

\paragraph{User prompt}
\begin{quote}
A tournament in the game ``Pre-Election Race'' is being conducted. The rules of a one-shot game of ``Pre-Election Race'' are given below:

1. Two candidates compete for the presidency of a fictional overseas country.

2. The overseas country has 9 states: A, B, C, D, E, F, G, H, I.

3. Each candidate has resources for 100 campaign trips.

4. Each of the two candidates simultaneously and independently decides how many times and to which states to travel. Each state may be visited any integer number of times from 0 to 100.

5. In each state, the candidate who visited that state more times wins. For winning each of the 9 states, the candidate receives 1 electoral vote. If the candidates visited a given state the same number of times, the election in that state ends in a tie, and both players receive 0.5 electoral votes.

6. The president is the candidate who receives more electoral votes. The winner receives 1 point. If the candidates receive the same number of electoral votes, each receives 0.5 points.

You are one of the candidates. Please indicate, for each of the nine states A, B, C, D, E, F, G, H, I, how many trips you will make to that state. In total, you may make no more than 100 trips.

All strategies entered into the tournament will play against one another in a round-robin format, that is, each strategy will play exactly one match against every other strategy. In the final tournament table, strategies are ranked by the total number of points earned across all presidential races.

In this tournament, your opponents will include both strategies submitted by other people and one strategy from each of several popular modern large language models. We expect several hundred human strategies and 5--10 strategies from different large language models. Your goal is to score as many points as possible across all presidential races and thereby finish as high as possible in the tournament standings.

In your answer, you must also provide a justification for your decision.

Required response format:

\texttt{\{"A": <number of trips>, "B": <number of trips>, "C": <number of trips>, "D": <number of trips>, "E": <number of trips>, "F": <number of trips>, "G": <number of trips>, "H": <number of trips>, "I": <number of trips>, "total": <sum of all trips, $\leq 100$>, "explanation": <why this allocation of trips was chosen>\}}
\end{quote}

\subsection{Tournament 3}

\paragraph{System message}
\begin{quote}\ttfamily
You are a tournament participant. Reply ONLY with a JSON object strictly following this schema:

\{"A": <int>, "B": <int>, "C": <int>, "D": <int>, "E": <int>, "F": <int>, "G": <int>, "H": <int>, "I": <int>, "total": <int $\leq$ 100>, "explanation": <string in Russian>\}

The total number of trips must be $\leq 100$. Do not add any text, any symbols outside the JSON object, or any Markdown.
\end{quote}

\paragraph{User prompt}
\begin{quote}
A tournament in the game ``Pre-Election Race'' is being conducted. The rules of a one-shot game of ``Pre-Election Race'' are given below:

1. Two candidates compete for the presidency of a fictional overseas country.

2. The overseas country has 9 states: A, B, C, D, E, F, G, H, I.

3. Each candidate has resources for 100 campaign trips.

4. Each of the two candidates simultaneously and independently decides how many times and to which states to travel. Each state may be visited any integer number of times from 0 to 100.

5. In each state, the candidate who visited that state more times wins. For winning each of the 9 states, the candidate receives 1 electoral vote. If the candidates visited a given state the same number of times, the election in that state ends in a tie, and both players receive 0.5 electoral votes.

6. The president is the candidate who receives more electoral votes. The winner receives 1 point. If the candidates receive the same number of electoral votes, each receives 0.5 points.

You are one of the candidates. Please indicate, for each of the nine states A, B, C, D, E, F, G, H, I, how many trips you will make to that state. In total, you may make no more than 100 trips.

All strategies entered into the tournament will play against one another in a round-robin format, that is, each strategy will play exactly one match against every other strategy. In the final tournament table, strategies are ranked by the total number of points earned across all presidential races.

In this tournament, your opponents will include both strategies submitted by other people and strategies from several popular modern large language models. We expect several hundred human strategies and approximately the same number of strategies from different large language models. Your goal is to score as many points as possible across all presidential races and thereby finish as high as possible in the tournament standings.

In your answer, you must also provide a justification for your decision.

Required response format:

\texttt{\{"A": <number of trips>, "B": <number of trips>, "C": <number of trips>, "D": <number of trips>, "E": <number of trips>, "F": <number of trips>, "G": <number of trips>, "H": <number of trips>, "I": <number of trips>, "total": <sum of all trips, $\leq 100$>, "explanation": <why this allocation of trips was chosen>\}}
\end{quote}

\end{document}